\documentstyle[12pt]{article}

0

\begin{document}
\hspace{12cm} hep-th/9612104
  
\vspace{20 mm}

\begin{center}

{\large \bf D-Brane Probes and Mirror Symmetry}

\end{center}

\vspace{10 mm}

\begin{center}{\bf C\'{e}sar G\'{o}mez}   

\vspace{7 mm}      

{\em Instituto de Matem\'{a}ticas y F\'{\i}sica Fundamental,CSIC, \protect \\ Serrano 123, 28006 Madrid, Spain}      

\vspace{5mm}             

\end{center}      

\vspace{25mm}

\begin{abstract} 
We study the effect of mirror symmetry for $K3$ surfaces on D-brane probe 
physics. The case of elliptically fibered $K3$ surfaces is considered in 
detail. In many cases, mirror can transform a singular fiber of Kodaira's 
type $ADE$ into sets of singular fibers of type $I_1$ ($II$) with equal 
total Euler number, but vanishing contribution to the Picard number of 
the mirror surface. This provides a geometric model of quantum splitting 
phenomena. Mirror for three dimensional gauge theories, interchanging 
Fayet-Iliopoulos and mass terms, is also briefly discussed.

\end{abstract}

\pagebreak


\section{Introduction.}

In this note we study mirror symmetry for $K3$ surfaces \cite{mirror}. We consider 
in detail the case of $K3$ surfaces which are elliptic fibrations. 
The physical implications of mirror transformations are studied using 
two and three D-brane probes \cite{Sf,BDS,Sir}, depending if we work M or F theory 
compactifications on $K3$. In the context of probe physics, we observe 
the following implications of mirror transformations. Given a 3D-probe \cite{Sf}
with Seiberg-Witten moduli \cite{SW} an elliptically fibered $K3$ surface, we observe 
that if we start with the $N\!=\!4$ classical solution, characterized by 
constant $\tau$, the mirror $K3$ elliptically fibered surface, describe 
the quantum corrected $N\!=\!2$ Seiberg-Witten solution with the mass 
parameters naturally appearing as moduli of the mirror $K3$ surface.In 
this context the splitting phenomena of a classical singularity into 
quantum singularities of the moduli appears as a result of mirror 
transformations on Shioda-Tate formula for elliptically fibered $K3$ 
surfaces with trivial group of sections \cite{CD,U}. The contribution of the ``classical'' 
singularity to the Picard lattice become part of moduli of the mirror surface 
where new type of singularities with no contribution to Picard should be 
introduced. Using the geometric mirror map \cite{SYZ,M} we observe that quantum corrections 
to the Coulomb branch of the probe dynamics can be mapped into ``internal'' 
instanton effects on the special lagrangian D-brain submanifold used to 
define the geometric mirror. To conclude we make some observations on the 
mirror pairs in $3$D recently discovered by Intriligator and Seiberg \cite{IS,BHOO,PZ,HW}, where 
the interchange between mass terms and Fayet-Iliopoulos terms become part 
of the mirror map on elliptically fibered $K3$ surfaces. An extended version of 
the work presented in this note is under preparation \cite{GH}.

\section{$K3$ surfaces}

A $K3$ surface is characterized by the Hodge diamond
\begin{equation} 
\begin{array}{ccccc}            
   &   & 1  &   &   \\    
   & 0 &    & 0 &   \\      
 1 &   & 20 &   & 1 \\         
   & 0 &    & 0 &   \\         
   &   & 1  &   &      
   \end{array}
\end{equation}
The space $H_2(X;{\bf Z}) \simeq {\bf Z}^{22}$ is a self dual 
lattice of signature $(3,19)$,
\begin{equation}
\Gamma_{3,19} = E_8 \perp E_8 \perp {\cal U} \perp {\cal U} \perp {\cal U},
\end{equation}
where ${\cal U}$, the hyperbolic plane, is the lattice ${\bf Z}^2$ with 
\( \left( \begin{array}{cc} 0 & 1 \\ 1 & 0                                                                             
\end{array} \right) \) and $E_8$ is the root lattice of $E_8$ with 
reversed sign,
\begin{equation} E_8 \equiv \left(     \begin{array}{ccccccccc}
  -2 &    &    & 1  &    &    &    &    \\                  
     & -2 &  1 &    &    &    &    &    \\                  
     &  1 & -2 & 1  &    &    &    &    \\                
   1 &    &  1 & -2 &  1 &    &    &    \\       
     &    &    &  1 & -2 &  1 &    &    \\       
     &    &    &    &  1 & -2 &  1 &    \\                                                                                                                                           
     &    &    &    &    &  1 & -2 &  1 \\                  
     &    &    &    &    &    &  1 & -2       
     \end{array} \right).
\end{equation}
A {\em marking} of the $K3$ surface is defined by an isomorphism of 
lattices,\begin{equation}\phi : H_2(X;{\bf Z}) \longrightarrow 
\Gamma_{3,19}.
\end{equation}
Given a complex structure we get the 
Hodge decomposition
\begin{equation}H^2(X;{\bf C}) = H^{0,2}(X) \oplus 
H^{1,1}(X) \oplus H^{2,0}(X).
\end{equation}  
Let $\Omega$ be a holomorphic 
$2$-form. The periods of the $K3$ surface are defined by
\begin{equation}
\overline{\omega_i} = \int _{e_i} \Omega,
\end{equation}
with $e_i$ a basis of $H_2(X;{\bf Z})$.  
Now we define the Picard lattice $\hbox{Pic} (X)$ by
\begin{equation}
\hbox{Pic}(X) \equiv H_{1,1}(X) \bigcap H_2(X;{\bf Z}).
\label{eq:7}
\end{equation}
$\hbox{Pic}(X)$ defines a sublattice $\Gamma_{1,t}$ of $\Gamma_{3,19}$. 
The rank of this lattice is given by
\begin{equation}\rho(X) = 1 + t.
\end{equation}
It is clear from (\ref{eq:7}) that $\rho(X) \leq 20$. The holomorphic 
$2$-form $\Omega \in H_2(X;{\bf Z})$ can be associated with a spacelike 
$2$-plane in ${\bf R}^{3,19}$:
\begin{equation}
\Omega = x +i y,
\end{equation}
with $x, y \in H_2(X;{\bf C}) \simeq {\bf R}^{3,19}$. From (\ref{eq:7}) it 
follows that the $2$-plane $\Omega$ is orthogonal to $\hbox{Pic}(X)$. 
  
The Teichm\"{u}ller space of complex structures \cite{A} is given by the grassmannian 
manifold of $2$-spacelike planes in ${\bf R}^{3,19}$:
\begin{equation}
{\cal T}_C = O(3,19)/O(2) \times O(1,19).
\end{equation}
Modding out by changes of marking we get the moduli space of complex structures
\begin{equation}
{\cal M}_C = O(3,19;{\bf Z}) \backslash O(3,19) / O(2) \times O(1,19).
\end{equation}
Generic changes of the complex structure will not preserve $\hbox{Pic}(X)$. The 
moduli of complex structures preserving $\hbox{Pic}(X)$ would be
\begin{equation}
O(\Lambda) \backslash O(2,19-t) / O(2) \times O(19-t),
\label{eq:12}
\end{equation}
where the {\em transcendental lattice} $\Lambda$ is defined by
\begin{equation}
\Lambda = \hbox{Pic}(X)^{\perp},
\end{equation}
the orthogonal lattice of $\hbox{Pic}(X)$ in $\Gamma_{3,19}$. Thus, $\Lambda$ 
is a lattice of type $(2,19-t)=(2,20-\rho(X)))$.

\section{Quantum Cohomology and Mirror Surfaces.}

Given $X$ with a Picard Lattice $\hbox{Pic}(X)$, we can try to find a mirror 
$K3$ surface $Y$ such that $\hbox{Pic}(Y)$ is given by the transcendental 
lattice of $X$. In order to do that, we need to extend the notion of 
$\hbox{Pic}(X)$ to that of {\em quantum} Picard $\Upsilon(X)$ \cite{A}. Let us define
\begin{equation}
\Upsilon(X) \equiv \hbox{Pic}(X) \perp {\cal U}.
\end{equation}
Thus, $\Upsilon(X)$ is a lattice of type $(2,t+1)=(2,\rho)$. In this sense, 
we can define the mirror $K3$ surface $Y$ through the condition
\begin{equation}
\Upsilon (Y) = \Lambda(X),
\end{equation}
with $\Lambda(X)$ the transcendental lattice of $X$. Notice that if 
$\hbox{Pic}(X)$ is of type $(1,t)$, then $\hbox{Pic}(Y)$ is of type 
$(1,18-t)$: 
\begin{equation}
\hbox{rank Pic}(X) + \hbox{rank Pic}(Y) =20. 
\end{equation}
The Teichm\"{u}ller space of complex structures of the mirror surface 
preserving $\hbox{Pic}(Y)$ is given by
\begin{equation}
{\cal T}_C(Y) = O(2,t+1)/O(2) \times O(t+1).
\end{equation}
The physical origin of mirror symmetry comes from the fact that
\begin{equation}
{\cal T}_{\sigma} = {\cal T}_C(X) \otimes {\cal T}_C(Y)
\end{equation}
is the Teichm\"{u}ller space for the $\sigma$-model defined on $K3$.
  
The mirror symmetry interchanging $X$ and $Y$ is now part of the modular 
group for the $\sigma$-model moduli,
\begin{equation}
{\cal M}_{\sigma} = {\cal T}_{\sigma} / {\cal M} \times {\cal M}_X \times {\cal M}_Y = 
O(4,20;{\bf Z}) \backslash O(4,20) / O(4) \times O(20), 
\label{eq:19}
\end{equation}
where {\cal M} in (\ref{eq:19}) simbolically represents the mirror 
symmetry trasnformation.
  
\vspace{2 mm}

{\bf Some examples}
  
\vspace{2 mm}

Notice that (\ref{eq:19}) is the Narain lattice for the heterotic string 
compactified on $T^4$ to $6$ dimensions. Moreover, (\ref{eq:19}) is also 
the moduli of type II$_A$ string compactified on $K3$. Recall type II$_A$ 
string contains RR sector with a one form and a three form. However, they 
do not contribute to the moduli, since $H_1=H_3=0$ for $K3$ surfaces.

We can now consider a case with $\hbox{Pic}(X)=\Gamma_{1,1}$. The moduli 
space (\ref{eq:12}) is given by 
\begin{equation}
O(2,18;{\bf Z}) \backslash O(2,18) / O(2) \times O(18),
\end{equation}
which is the Narain lattice for heterotic string on $T^2$. This is a 
result known as duality between $F$-theory \cite{V} on $K3$, with $\hbox{Pic}(K3)=
\Gamma_{1,1}$ and heterotic string on $T^2$ \cite{MV}.
  
As another example, we can consider $M$-theory on $K3$: we must take the 
moduli of $M$-theory on $K3$ to be 
\begin{equation}
O(3,19;{\bf Z}) \backslash O(3,19) / O(3) \times O(19),
\end{equation}
that corresponds to heterotic string on $T^3$.
  
In a certain sense we observe that M and F theories are different ways of 
mapping $K3$ moduli into heterotic moduli.

\section{Polarized $K3$ Surfaces.}

Following Dolgachev \cite{D}, we define an $M$-polarized $K3$ surface as a pair 
$(X,j)$, with $j$ a lattice embedding,
\begin{equation}
j:M \longrightarrow \hbox{Pic}(X),
\end{equation}
such that
\begin{equation}
\phi^{-1}(M) \in \hbox{Pic}(X).
\end{equation}
We will take $M$ to be a lattice of type $(1,t)$. As before, the moduli of 
complex structures of $M$-polarized $K3$ surfaces would be given by 
(\ref{eq:12}), with $t+1$ the rank of $M$. We will take 
\begin{equation}
\hbox{rank } M = \hbox{rank Pic}(X). 
\end{equation}
To define the mirror surface we choose an isotropic vector $f$, with 
$(f,f)=0$, in $M^{\perp}$, the orthogonal to $M$ in $\Gamma_{3,19}$. 
Defining
\begin{equation}
M^* = f^{\perp}/f,
\end{equation}
the mirror $K3$ surface is the $M^*$-polarized $K3$ surface. Notice that if 
$M$ is of type $(1,t)$, then $M^*$ is of type $(1,18-t)$, as required by 
mirror. As an example, consider
\begin{equation}
M = < 2n >, 
\label{eq:27}
\end{equation}
the lattice defined by $e \cdot e = 2n$. The orthogonal $M^{\perp}$ is given by
\begin{equation}
M^{\perp} = {\cal U} \perp {\cal U} \perp E_8 \perp E_8 \perp < -2n >,
\end{equation}
and $M^*$,
\begin{equation}
M^{*} = {\cal U} \perp E_8 \perp E_8 \perp < -2n >.
\label{eq:29}
\end{equation}
Notice that choosing $f$ in this construction is equivalent to finding a 
``classical'' Picard sublattice in $M^{\perp}$. The example (\ref{eq:27}) 
corresponds to $t=0$, and $M^*$ in (\ref{eq:29}) is of $\hbox{rank}=19$.

\section{Elliptic Fibrations.}

A $K3$ surface $X$ is an elliptic fibration if 
\begin{equation}
\Pi : \: \:  X \longrightarrow {\bf P}^1
\end{equation}
with $\Pi^{-1}(z)$ an elliptic curve. The basic information on an elliptic fibration 
is given by its degenerate fibers. They were classified by Kodaira (see Table 1) \cite{K}.
  
Denoting $F_v$ the set of singular fibers, we have
\begin{equation}
24 = \sum e(F_v),
\label{eq:31}
\end{equation}
with $e(F_v)$ the Euler number of the singular fibers. An equivalent 
condition to (\ref{eq:31}) is given by the adjunction formula \cite{GHpag},
\begin{equation}
K_x = \Pi^* ( K_{{\bf P}^1} + \sum a_i P_i),
\end{equation}
with $P_i$ the points on the base space where the fiber becomes singular, 
and $a_i$ given in Table 1.
  
From Shioda-Tate lemma the following formula can be derived: 
\begin{equation}
\rho(X) = 2 + \sum_{v} \sigma(F_v) + \hbox{rank } \Phi,
\label{eq:33}
\end{equation}
where $\rho(X)$ is the Picard number of $X$, $\Phi$ the group of sections and 
$\sigma(F_v) + 1$ the number of components of the singular fiber $F_v$. In 
Kodaira's classification this number is given by the number of points 
of the affine Dynkin diagram (see Table 1).

\begin{center}

\begin{tabular}{|c|c|c|c|c|}     \hline\hline
        {\bf Kodaira} & {\bf Dynkin}      & ${\bf e}$ & ${\bf \sigma}$ & {\bf a} \\ \hline 
        $I_n$         & $\tilde{A}_{n-1}$ & $n$       & $n-1$          & $n/12$  \\ \hline
        $I_{n}^*$     & $\tilde{D}_{4+n}$ & $n+6$     & $n+4$          & $1/2 + n/12$  \\ \hline
        $II$          &                   & $2$       & $0$            & $1/6$  \\ \hline
        $III$         &                   & $3$       & $1$            & $1/4$  \\ \hline
        $IV$          &                   & $4$       & $2$            & $1/3$  \\ \hline
        $II^*$        & $\tilde{E}_{8}$   & $10$      & $8$            & $5/6$  \\ \hline
        $III^*$       & $\tilde{E}_{7}$   & $9$       & $7$            & $3/4$  \\ \hline
        $IV^*$        & $\tilde{E}_{6}$   & $8$       & $6$            & $2/3$  \\ \hline\hline
\end{tabular}

\end{center}
  
The meaning of the Shioda-Tate lemma is the following. Given the elliptic fibration, 
we can define $\hbox{Pic'}(X)$ as the Picard sublattice containing a fiber 
$F$ and a section $S$ satisfying
\begin{eqnarray}
F \cdot S & = & 1,   \nonumber \\ 
S \cdot S & = & -2,  \nonumber \\ 
F \cdot F & = & 0, 
\label{eq:34}
\end{eqnarray}
and the sum of lattices of type $ADE$ spanned by the irreducible components 
of fibers not intersecting the section $S$. The rank of $\hbox{Pic'}(X)$ 
is given by 
\begin{equation}
\rho ' (X) = 2 + \sum_v \sigma (F_v).
\end{equation}
Thus, equation (\ref{eq:33}) becomes equivalent to the isomorphism 
between $\hbox{Pic}(X) / \hbox{Pic '}(X)$ and the group of sections of the fibration.
  
We can now consider some examples. Let us consider $M=<+2>$; the mirror 
manifold is defined by\footnote{As shown 
by Dolgachev \cite{D}, (\ref{eq:36}) defines the mirror \cite{GP,B} to the toric manifold 
defined by $(1,1,1,3)$, $x_1^6+x_2^6x_3^6+x_4^2=0$. This $K3$ surface 
is specially interesting, since it is the one appearing in the $K3$-fibration 
of the Calabi-Yau space ${\bf P}^{12}_{\{1,1,2,2,6\}}$, used to obtain 
Seiberg-Witten moduli from heterotic-type II dual pairs \cite{KKLMV,GHL}.}
\begin{equation}
M^* = {\cal U} \perp E_8 \perp E_8 \perp <-2>.
\label{eq:36}
\end{equation}
This lattice is of $\hbox{rank }=19$. Let us now consider this lattice as 
$\hbox{Pic '}(X)$ with a trivial group of sections. Using Table $1$ and 
equation (\ref{eq:33}), we can define the mirror as an elliptic fibration 
with two fibers of type $E_8$ and a fiber with $\sigma=1$, i.e., either of type 
$I_2$ or $III$. In order to saturate (\ref{eq:31}), we need to add singular fibers 
of type $I_1$, i.e., with $\sigma=0$.
  
Using the same type of arguments we can consider some pairs of mirror $K3$ 
surfaces which are both elliptically fibered. Let us consider for instance a case 
with $\rho(X)=2$, and $24$ singular fibers of type $I_1$. In this case, 
the Picard lattice $\Gamma_{1,1}$ would be interpreted as generated by a 
section $S$ and a fiber $F$ staisfying relations (\ref{eq:34}). The 
mirror to this $K3$ surface would be given by
\begin{equation}
M = {\cal U} \perp E_8 \perp E_8,
\end{equation}
with Picard number $18$. This corresponds to an elliptic fibration with two 
$E_8$ singularities and extra $I_1$'s not contributing in (\ref{eq:33}) to 
the Picard number $\rho(X)$. This is the elliptic fibration used in F-theory 
to define the type II$_B$ dual of heterotic string on $T^2$, with 
unbroken $E_8 \times E_8$ symmetry, i.e., no Wilson lines \cite{MV}.

\section{Mirror Symmetry and Sen's Orientifold Model.}

In \cite{Sf} Sen has considered an elliptically 
fibered $K3$ consisting of four $D_{4}$ singularities. This corresponds 
to constant $\tau$, since for $D_{n}$ singularities the monodromy is 
given by
\begin{equation}
\left( \begin{array}{cc} -1 & -n+4  \\
                          0 & -1
       \end{array} \right).
\end{equation}
From Shioda-Tate lemma and assuming a trivial group of sections we get
\begin{equation}
\rho = 18
\label{eq:38}
\end{equation}
as the Picard number. The situation corresponding to a $D_{4}$ singularity 
can be described in terms of a type $IIB$ compactification with four 
orientifold planes, contributing with charge $-4$, and four groups of 
$7$-branes compensating locally the charge of the orientifold \cite{Sf}. This 
explains in type $IIB$ language the constant value of $\tau$. Equivalently, 
in F-theory language each $D_{4}$ singularity contributes with six units to 
the total Euler number. Let us now consider the mirror to the model defined 
by four $D_{4}$ singularities. This is a $K3$ surface with Picard lattice of 
type $\Gamma_{1,1}$. The dimension of the moduli space for the mirror 
surface is given by the Picard number (\ref{eq:38}), i.e., $18$. The singular 
fibers for the mirror surface should be of type $I_{1}$ or $II$ in Kodaira's 
notation. In fact they contribute to the total Euler number, but not 
to the Picard number of the mirror $K3$ surface, which is equal to two. 
The $18$ moduli of the mirror can be described using four $SU(2)$ 
$N_{f}\!=\!4$ SUSY gauge theories, each one describing locally a ``quarter'' 
of the base space \cite{Sf}. For each of these theories we have four moduli 
corresponding to the masses of the $N_{f}\!=\!4$ hypermultiplets. 
The other two common moduli, up to the total of 18, correspond to 
the complex and K\"{a}hler moduli of the heterotic dual on $T^{2}$. The 
moduli $\tau_{0}$ in Seiberg-Witten solution for $SU(2)$ with 
$N_{f}\!=\!4$ is related to the ratio, fixed by the $j$-function, of 
these two moduli. 
   
   The interest of Sen's result is in part due to the 
observation that the F-theory solution contains the quantum corrections, 
which are at the origin of the orientifold splitting. Our previous analysis 
in terms of mirror $K3$ surfaces suggest the following general picture: In 
going to the mirror surface we replace the $D_{4}$ singularity by a set of 
singularities with total Euler number equal to $6$, but not contributing 
to the Picard number. Simbolically, 
we can define the mirror action on the singular fiberes 
as follows\footnote{Other possibilities for the quantum 
splitting (\ref{eq:39b}) are \(D_4 \rightarrow \oplus_{i=1}^{3} II^{i}; \: 
D_4 \rightarrow II \oplus II \oplus + I_1 \oplus I_1; \: D_4 \rightarrow 
II + \oplus _{i=1}^{3} I_1^{i} \).}:
\begin{eqnarray}
M : D_4         & \longrightarrow & \oplus_{i=1}^{6}I_1^{i}, \nonumber \\
M : \sigma(D_4) & \longrightarrow & \sigma(\oplus_{i=1}^{6}I_1^{i})=0.
\label{eq:39b}
\end{eqnarray}
Generically, and for trivial group of sections, for any singular fiber $S$
of $ADE$ type, the ``mirror'' $S^*$ would satisfy 
\begin{eqnarray}
     e(S^*) & = & e(S), \nonumber \\
\sigma(S^*) & = & 0. 
\end{eqnarray}

By this process we pass from the moduli of the 
original theory described by the $D_{4}$ singularities to the $18$ 
dimensional moduli of the mirror manifold. The Seiberg-Witten theory 
used by Sen, i.e., the quantum moduli of the $3$-brane probe world volume 
field theory, is parametrized by these moduli. In summary we observe that 
quantum corrections are automatically encoded in the $K3$ mirror surface, 
i.e., the moduli of the mirror parametrize the quantum deformations. 
Moreover the splitting phenomena can be also interpreted as a mirror 
effect that replaces a singularity contributing to the Picard number 
by a set of singularities with equal Euler number but not contributing, 
in the Shioda-Tate formula, to the total Picard number of the mirror 
manifold\footnote{The use of mirror symmetry to derive quantum moduli for 
supersymmetric gauge theories is also considered in reference \cite{KKV}. It 
would be interesting to compare both approaches.}.

\section{Geometric Mirror and D-Branes}

In reference \cite{SYZ} a geometric characterization of mirror manifolds was 
propposed using toroidal lagrangian submanifolds. In this approach, 
mirror becomes equivalent to T-duality. Again, we will reduce our 
analysis to the simpler case of $K3$ surfaces. A special lagrangian 
submanifold in a $K3$ surface $X$ is a compact complex one dimensional 
manifold $M$ with an inmersion $f: M \rightarrow X$, such that 
$f^*(\Omega)$ coincides with the induced volume form on $M$. The 
moduli of special lagrangian submanifolds is determined by Mc Lean's 
theorem. Its dimension is equal to $b_{1}(M)$. We will take for M the 
$1$-torus $T$ with $b_{1}(T)\!=\!1$. Denoting $M(T,X)$ the moduli of 
immersions, the space $X$ becomes an elliptic fibration with base space 
$M(T,X)$. Now, we consider a $U(1)$ flat bundle on $T$, and define 
the D-brane moduli space $M_{D}(T,X)$ containing the moduli of immersions 
and the moduli of flat $U(1)$ bundles. The space $M_{D}(T,X)$ fibers on 
$M(T,X)$. The geometric mirror is the statement that $M_{D}(T,X)$ is the 
mirror $K3$ surface\footnote{It is important to notice the 
similarity between $M_D(T,X)$ and Donagi-Witten \cite{DW} construction of 
integrable models. For the hyperk\"{a}hler case, i. e., $K3$ surfaces, 
we can use to describe the moduli space of immersions of $T$ in 
$X$ the holomorphic one forms on $T$. In this sense, and interpreting $T$ 
as the reference surface $E_{\tau}$ in \cite{DW}, the Higgs field 
$\phi$ on $E_{\tau}$ can be interpreted as parametrizing the 
different immersions on $K3$.}. Thus in this approach 
we pass from an elliptic fibration to a mirror surface which is also 
elliptically fibered. Based on Mukai's results \cite{Mu}, Morrison \cite{M} argues that, 
at least for $K3$ surfaces, geometric mirror coincides with the standard 
concept of mirror above described. Geometric mirror provides a different 
point of view on the dynamical origin of the quantum corrections originating 
the splitting phenomena. Coming back to Sen's case, if the "mirror" of 
the $D_{4}$ singularity is described by six $I_{1}$ singularities and at 
the same time this mirror elliptic fibration is interpreted as $M_{D}(T,X)$, 
then we observe that the dynamics underlying the "splitting" of $D_{4}$ into 
$I_{1}$'s is due to instanton effects on the D-brane $T$, that we should 
describe as a disc in $X$ winding on one-cycles of $T$ \cite{SYZ}. In this way, 
"internal manifold" instanton effects allow us to pass from the "classical" 
$\tau$ constant solution to the quantum "mirror" Seiberg-Witten solution.

\vspace{2 mm}

{\bf T-Duality and Probes}
  
\vspace{2 mm}

The equivalence between geometric mirror and T-duality, can be 
simbolically represented as follows:
\begin{equation}
{\cal M}(\bullet ;X^*) = {\cal M}_D(T;X),
\label{eq:41}
\end{equation}
where ${\cal M}(\bullet,X^*)$ represents the mirror of a $0$-brane on the $K3$ 
surface $X^*$, and ${\cal M}_D(T;X)$ the D-brane moduli on the mirror manifold\footnote{The map 
from $2$-cycles into $0$-cycles is part of the Fourier-Mukai transform \cite{BBH} for 
$K3$}. 
Notice that for $K3$ surfaces mirror is duality in the sense that
\begin{equation}
X^{**}=X.
\end{equation}
  
Now we can consider the case of M-theory on $K3$ and consider a $2$-brane 
probe defining a $3$D SUSY theory on its worldvolume. The moduli of this 
$3$D SUSY theory is given by $K3$ itself. This $2$-brane looks from the 
$K3$ point of view as a $0$-brane, and therefore we can identify the left 
hand side of (\ref{eq:41}) with the moduli of the $3$D SUSY theory defined 
on the $2$-brane worldvolume \cite{SW3d,GMS}. Now, how should we interpret (\ref{eq:41}) 
in terms of pobe dynamics?
  
We can try to interpret the right hand side of (\ref{eq:41}) as the moduli of 
a $4$-brane with worldvolume space ${\bf R}^2 \times S^1 \times S^1$, i. e., 
a $5$D gauge theory. In this case, the moduli of this $5$D \cite{S5d} would be given 
by the base space of the fibration ${\cal M}_D(T;K3)$, i. e., the Mc Lean space 
of deformations of inmersions of $T$ in $K3$. When we compactify on 
$T$ to three dimensions we get the moduli of the $3$D theory of the 
$2$-brane probe on the {\em mirror surface}.
  
Moreover, after compactification from $5$D to $3$D on $T^2$ the theory on the 
$2$-brane probe posseses $N\!=\!4$ matter content. Again, we find the same picture as in Sen's example. 
The case with $N\!=\!4$ goes to the deformed $N\!=\!2$ in passing to the 
mirror surface.
  
Another specially interesting case of geometric mirror is the one of 
Calabi-Yau fourfolds, where the special lagrangian manifold is a complex four 
torus. The results of reference \cite{BJPSV} concerning Seiberg's 
duality \cite{Sd} for $N\!=\!1$ gauge theories can be interpreted 
as a consequence of mirror symmetry on the fourfold, i. e., $T$-duality 
transformations on the special toroidal lagrangian. This example indicates 
that F-theory compactifications on mirror manifolds produce 
dual field theories on the worldvolume probe.

\section{Mirror Symmetry in $3$D}

For $3$D gauge theories a type of mirror has been recently discovered by 
Intriligator and Seiberg \cite{IS} using the D-brane probe philosophy. Mirror 
symmetry in this case interchanges the Coulomb and Higgs branches. If 
the Coulomb branch is a $K3$ surface with $ADE$ singularities, the Higgs 
branch is defined as the corresponding moduli of $ADE$ instantons. Notice 
that the $K3$ is elliptically fibered and that in the D-probe philosophy 
singular fibers define the global symmetries for the $2$-brane probe 
worldvolume physics. Im more concrete terms, mirror symmetry discussed in 
\cite{IS} interchanges masses, on which the Coulomb branch geometry 
depends quantum mechanically, and Fayet-Iliopoulos D-terms, which can 
only affect the metric of the Higgs branch. This is symmetry consistent 
with our approach in previous sections to mirror symmetry in the following 
sense: the number of Fayet-Iliopoulos terms is precisely given by the value 
of $\sigma$ for the singularity (see Table 1). This is the contribution to 
the Picard lattice in Shioda-Tate formula of the corresponding singularity. 
Generically, as we pass to the mirror $K3$ surface we convert this contribution 
to Picard into moduli parameters of the mirror surface. 
These moduli parameters, as was described in Sen's example, are interpreted 
as mass terms on which the metric of the Coulomb branch depends. In this sense, 
the interchange between Fayet-Iliopoulos and mass terms reflects the 
transformation, by mirror symmetry, of the Picard number of the 
singular fibers into moduli of the mirror surface.
  
Notice that for Kronheimer $ADE$ gauge theories, the number of mass 
parameters for $A_{n-1}$ theories is one and zero for $D,E_6,E_7,E_8$ 
theories. In our approach, the mass term for $A_{n-1}$ theories comes 
from the fact that for $I_n$ singularities $e(I_n) = \sigma(I_n) + 1$, 
while for $D,E_6,E_7,E_8$ we have $e(\: \:) = \sigma(\: \:) + 2$.

\vspace{10 mm}
  
\begin{center} 

{\bf Acknowledgements}

\end{center}

\vspace{5 mm}

I wish to thank U. Bruzzo for useful discussions, and the theory group at 
SISSA, where this work was partially done. This research was partially 
supported by grant AEN96-1655 and by European Community grant ERBCHBGCT 94 06 34.

\end{document}